\documentclass[12pt]{article}
\usepackage{graphicx}
\usepackage{amssymb,amsmath,amsfonts,palatino,amsthm}
\usepackage{amssymb}
\usepackage{epstopdf}
\newcommand{\f}{\begin{equation}}
\newcommand{\ff}{\end{equation}}

\begin{document}

\title{Precedence and freedom in quantum physics \\}
\author{Lee Smolin\thanks{lsmolin@perimeterinstitute.ca} 
\\
\\
Perimeter Institute for Theoretical Physics,\\
31 Caroline Street North, Waterloo, Ontario N2J 2Y5, Canada}
\date{\today}
\maketitle

\begin{abstract}

A new interpretation of quantum mechanics is proposed according to which precedence, freedom and novelty play central roles.  This is based on a modification
of the postulates for quantum theory given by Masanes and Muller\cite{MM}.  We argue that quantum mechanics is uniquely characterized as the probabilistic theory in which individual systems have maximal freedom in their responses to experiment, given reasonable axioms for the behaviour of probabilities in a physical theory.  Thus, to the extent that quantum systems are free, in the sense of Conway and Kochen\cite{CK}, there is a sense in which they are maximally free. 

We also propose that laws of quantum evolution arise from a {\it principle of precedence} according to which the outcome of a measurement on a quantum system is selected randomly from the ensemble of outcomes of previous instances of the same measurement on the same quantum system.  This implies that dynamical laws for quantum systems can evolve as the universe evolves, because new precedents are generated by the formation of new entangled states.  

\end{abstract}

\newpage

\tableofcontents

\section{Introduction}

We are used to thinking that the laws of physics are deterministic and that this precludes the occurrence of genuine novelty in the universe.  All that happens is rearrangements of elementary particles with unchanging properties by unchanging laws.  


But must this really be the case?  We need determinism only in a limited set of circumstances, which is where an experiment has been repeated many times.  In these cases we have learned that it is reliable to predict that when we repeat an experiment in the future,  which we have done many times in the past, the probability distribution of future outcomes will be the same as observed in the past.  

Usually we take this to be explained by the existence of fundamental timeless laws which control all change.  But this could be an over-interpretation of the evidence.  What we need is only that there be a principle that measurements which repeat processes which have taken place many times in the past yield the same  outcomes as were seen in the past.  Such a {\it principle of precedence} would explain all the instances where determinism by laws works without restricting novel processes to yield predictable outcomes.  There could be at least a small element of freedom in the evolution of novel states without contradicting the application of laws to states which have been produced plentifully in the past.

But are there any truly novel states in nature?

It is fair to say that classical mechanics precludes the existence of genuine novelty, because for certain all that happens is the motion of particles under fixed laws.  But quantum mechanics is different, in two ways.  First, in quantum mechanics does not give unique predictions for how the future will resemble the past. It gives from past instances only a statistical distribution of possible outcomes of future measurements.

Second, in quantum physics there is the phenomena of entanglement which involves novel properties shared between subsystems which are not just properties of the individual subsystems. The free will theorem of Conway and Kochen\cite{CK} tells us that in these cases systems respond to measurements in a way that can be considered free, in the sense that the result of an individual measurement on elements of an entangled system could not be predicted by any knowledge of the past.  

 An entangled state can be novel in that it can be formed from a composition of subsystems into a state never before occurring in the prior history of the universe.  This is common for example in biology where natural selection can give rise to novel proteins and sequences of nucleic acids which almost certainly, due to the combinatorial vastness of the number of possibilities, have not existed before.

There is then the possibility that novel states can behave unpredictability because they are without precedent. Only after they have been created enough times to accumulate ample precedent would the behaviour of these novel states become lawful.

Hence we can have a conception of law which is sufficient to account for the repeatability of experiments, without restricting novel states from being free from constraints from deterministic laws.  In essence the laws evolve with the states.  The first several iterations of a novel state are not determined by any law.  Only after sufficient precedent has been established does a law take hold, and only for statistical predictions.   Individual outcomes can be largely unconstrained.  

Quantum physics allows this possibility because the generic single measurement is not determined by quantum dynamics.  Only if the system is prepared in an eigenstate of the measurement being made is the result determined.  But these require fine tuning and are hence non-generic.   Otherwise quantum dynamics is stochastic so that no outcome of a single generic observation can disagree with predictions of quantum mechanics.  


There are aspects of measurements that are not predicted by quantum mechanics which offer scope for genuine novelty and freedom from deterministic evolution.  Imagine a double slit experiment with a very slow source of photons.  The measurement gives a sequence of positions to which the photons fall on the screen,  $x_1, x_2,  . . .  ,  x_N$.  Each individual photon can end up anywhere on the screen.  Quantum mechanics predicts the overall statistical ensemble that accumulates after many photons, $\rho (x)$.  But it does not, for example, restrict the order by which they fall.  Quantum mechanics is equally consistent with a record in which the $x_i's$ are permuted, from one random sequence to another. 

Macroscopic outcomes could depend on the order of positions, for example, if someone chooses to make a career in science or politics based on whether the $13$'th photon falls to the left or right side of the screen. 

The basic idea of the formulation of quantum theory proposed here is that 1) systems with no precedents have outcomes not determined by prior law and 2) when there is sufficient precedence the outcome of an experiment is determined by making a random selection from the ensemble of prior cases. 3) The outcome of measurements on systems with no or few precedents is as free as possible, in a sense that needs to be defined precisely.  Stated more carefully these become the principles of this approach to quantum theory, to be enunciated below. 

This proposal is a twist on the real ensemble interpretation proposed earlier in \cite{real1}.  The principle proposed there was that whenever probabilities appear in quantum physics they must be relative frequencies within ensembles every element of which really exists.  In the original version of this idea the ensemble associated with a quantum state exist simultaneously with it.  In the current version the ensembles exist in the past of the process they influence.  

How much precedence is necessary to turn freedom into deterministic dynamics?  There must for each system be an answer to this question.  

If the first instance of a measurement made on a novel state is undetermined, but the probabilities for outcomes of a measurement with a great deal of precedence is tightly determined, there is, for any system, a number of distinct  prior preparations whose statistical distributions of outcomes must be measured to determine, as well as can be done, the distributions of  outcomes of  measurements made on future iterations of that system.  This is the number of degrees of freedom of the system, to be denoted $K$ below\cite{Lucien,MM}.  There is also the dimension or capacity of the system which is the number, $N$, of outcomes that can be distinguished by measurements on the system\cite{Lucien,MM}.  These numbers and their relation must play a crucial role because they determine when there is sufficient precedent for future cases to be determined as possible.

We show below that there is a precise sense in which quantum kinematics is specified by requiring that $K$ be as large as possible, given $N$, consistent with a small set of reasonable general axioms.  This means that there is the maximal amount of information needed per distinguishable outcome to predict the statistical distribution of outcomes for any experiment.  As a result, we can say  that the responses of quantum systems  to individual measurements are maximally free from the constraints of determinism from prior cases. 


To formulate this idea precisely, we can make use of an axiomatic formulation of quantum theory, given by  Masanes and Muller\cite{MM}.  (The idea of formulating quantum mechanics in terms of simple operational axioms was introduced by Hardy\cite{Lucien}).  They give four axioms for how probabilities for outcomes behave when systems are combined into composite systems, or subsystems are projected out of larger systems and proves that they imply quantum mechanics or classical probability theory. To these we add a new, fifth,  axiom which pick out the quantum case.   These five postulates define the kinematics of quantum systems.  

The hard work needed to show this has already been done by Masanes and Muller\cite{MM}, my observation that these five postulates determine quantum theory is a trivial consequence of their work.

Informally stated these five postulates are

\begin{enumerate}

\item{}The state of a composite system is characterized by the statistics of measurements on the individual components.

\item{}All systems that effectively carry the same amount of information have equivalent state spaces.

\item{}Every pure state of a system can be transformed into every other by a reversible transformation.

\item{}In systems that carry one bit of information, all measurements which give non-negative probabilities are allowed by the theory.

\item{}Quantum systems are maximally free, in that a specification of their statistical state, sufficient for predicting the probabilities for outcome of any future 
measurement,  requires the maximal amount of information, relative to the number of outcomes of an individual measurement. 

\end{enumerate}

To these we add a postulate about quantum dynamics.  This is the principle of precedence, which,  informally stated,  says

{\bf Principle of precedence}:  When a quantum process terminating in a measurement has many precedents, which are instances where an identically prepared system was subject to the same measurement in the past, the outcome of the present measurement is determined by picking randomly from the ensemble of precedents of that measurement.

We give now a brief sketch of this novel interpretation of quantum mechanics, by giving more precise statements of these postulates.

\section{New postulates for quantum theory}

We give two sets of postulates, the first, which pick out the kinematical framework and the second, which specifies dynamics.  

\subsection{Definitions}

We need a number of definitions, in most cases more information is found from \cite{Lucien,MM}, from where these definitions are taken. 

\begin{itemize}

\item{} The universe at a given time consists of a number of systems, $S_I$.

\item{} Systems are labeled by their constituents and preparation.

\end{itemize}

New systems can be formed from old systems in three ways.

\begin{itemize}

\item{}New systems can be formed by combinations of existing systems, or composition.  
\f
S_{12}= S_1 \cup S_2
\ff

\item{}Systems may also be prepared by projection, which is subjecting them to a filter that picks out a subset of possible measurement outcomes.

\item{}Systems may also be altered by evolution in time, with or without external influences.  This is also called transformation.  

\end{itemize}

Composition increases the set of possible outcomes of measurements, projection reduces it and evolution leaves the number fixed.  

\begin{itemize}

\item{}{\bf Capacity:} For a given system, there is a number $N$, called the capacity by Masanes and Muller\cite{MM}, which  is the number of possible outcomes of a given single measurement made on the system. 

\item{}{\bf Degrees of freedom:} The minimal amount of information needed to completely determine the statistical distribution of outcomes of any experiment on the system is $K$ real parameters\cite{Lucien,MM}.  

\item{}{\bf Statistical state} The state of a system $S$ can be specified by a list of $K$ probabilities, $ \rho= (p_1,  . . .  ,p_K)$ which are complete in the sense that the probabilities of any measurement made on $S$ can be computed from the $p_a$.   The space of states, ${\cal S}_S$ is convex and compact, because the probabilities
are bounded by $0 \geq p_a \geq 1$.  A state is a mixture if it can be written as a statistical mixture of two other states,
\f
p_a = x p^1_a + (1-x) p^2_a 
\ff
for a probability $0 \geq x \geq 1$.  A state is pure if it cannot be so expressed.

\item{}{\bf Measurement} is a map $E: \rho \rightarrow R$ which corresponds to a possible experiment.  A set of complete measurements is $N$ measurements that suffice to pick out uniquely one of the $N$ possible outcomes of a measurement.

\end{itemize}

\subsection{Kinematical postulates for quantum theory}

For the kinematical framework we work in the tradition of operational axioms for quantum theory pioneered by Hardy.  I find it most useful to use a 
set of postulates proposed by Masanes and Muller\cite{MM}.

Masanes and Muller give four postulates for quantum theory.  To emphasize the role of freedom in quantum physics I would like to propose a modification of their postulates which modify one and add one additional, so that we have the following system of five postulates.

\begin{itemize}

\item{}{\bf Postulate 1: Local tomography}. The state of a composite system $AB$ is completely characterized by the statistics of measurements on the subsystems $A$ and $B$.

\item{}{\bf Postulate 2: Equivalence of subspaces.} Let $S_N$ and $S_{N-1}$ be systems with capacities $N$ and $N - 1$, respectively. If $E_1, . . . , E_N$ is a complete measurement on $S_N$ , then the set of states $\omega \in S_N$ with $E_N (\omega ) = 0$ is equivalent to $S_{N-1}.$  Physically this means that if two quantum systems of the same capacity are equivalent even if one arises by reduction from a larger system.  

\item{}{\bf Postulate 3: Symmetry}. For every pair of pure states $\omega, \phi \in S_A $  there is a reversible transformation 
$T $ such that $T \cdot \omega =  \phi $. 

\item{}{\bf Postulate 4: All measurements allowed.} All probability measures on $S_2$ (that is, maps from $S_2$ to the interval $[0,1]$ are outcome probabilities of possible measurements.

\item{}{\bf Postulate 5: The principle of maximal freedom}. The amount of  information needed to predict the statistical distribution of outcomes of any experiment on a system $S$ should be as large as possible, given $N$.  Thus, $K$ should increase with $N$ by a fixed function, which grows as fast as possible, consistent with the other axioms.  

\end{itemize}

We note that Postulates 1 to 4 are given by 
Masanes and Muller\cite{MM}.  Note that we use a version of Postulate 3 that does {\bf not} specify the transformations are continuous.  The only thing new is the fifth postulate.  The fact that these five postulates together pick out quantum theory uniquely follows trivially from Theorem 1 of Masanes and Muller which asserts that the first four postulates have two realizations, classical probability theory and quantum theory.  Postulate 5 then trivially picks out quantum theory, for which $K =N^2 -1$ over classical probability theory for which $K=N-1$\cite{Lucien,MM}.

\subsection{Dynamical postulates for quantum theory}

We begin by postulating a real ensemble within which probabilities for quantum systems are defined as relative frequencies.    

\begin{itemize}

\item{}{\bf Definition:} The {\it precedents } of a  quantum system $S$ is the ensemble, $E(S)$ of systems with the same constituents and preparation (including transformations) in the past.  The ensemble of {\it precedents of  a measurement}, $M$, consists of  copies of the processes with the same constituents, preparation and measurement in the past.  $M(E,S)$ is the ensemble of outcomes of these measurements. 

\end{itemize}

We also need to specify how the formalism applies to experiment.  This is through a 

\begin{itemize}

\item{}{\bf Principle of correspondence}: The statistical state $\rho$ of a quantum system, $S$, is a description of the ensemble of its precedents. It is measured
by constructing an ensemble and measuring the probability distributions of the outcomes of $K$ distinct experiments on them.

\end{itemize}

Now we are ready to state the dynamical principle of quantum physics. 

\begin{itemize}

\item{}{\bf Postulate 6: Precedence:}  The outcome of a measurement, $M$, on a system, $S$, is a randomly chosen member of $M(E,S)$, the ensemble of outcomes of past instances of that measurement on identically prepared systems, in the case that the number of such precedents is large.

\end{itemize}

What if a system has no precedents? This can occur if an entangled state is formed for the first time. 
Then it is a novel state and we apply

\begin{itemize}

\item{}{\bf The principle of freedom in the absence of precedent:} A quantum system, $S,$  may have no precedent.   Then the outcome of a measurement $M$ on it is not determined by any prior knowledge of the state of the universe.  

\end{itemize}

A system may have aspects that have precedence without being determined totally.  For example, the possible outcomes of a measurement on a novel system are still constrained by symmetries.  They are also constrained by the conservation laws that follow from symmetries including energy.  

What happens in between, when the number of precedents is non-zero but small.  This requires a novel principle, about which I only have a few preliminary remarks to offer in the next section.  A question we can answer now is how many instances are necessary to go between the principle of freedom and the principle of precedence?  We note that $K$ is a measure of how many precedents are necessary before the statistical distributions of the outcome of any experiment are determined.  Hence, $K$ is a measure of the freedom of quantum systems.  This is specified by postulate 5.

This means that quantum systems are maximally free because a maximal number of prior cases is needed to establish enough precedence to predict as we as can be done the result of any measurement.  To put this another way, it takes a maximal amount of information to be able to predict or foresee how the system will respond to anything it might encounter.  

 The fact that $K$ is large compared to $N$ reflects the fact that a density matrix, which could arise as the description of the state of a subsystem entangled within a larger system, requires much more information to specify, then a pure state.    This means that there are many more ways a system may be entangled within a larger system than can be distinguished by a single measurement.  This accords a great freedom to quantum systems because it means that they can have properties that are only expressed by measuring statistical distributions over many repeated experiments.  Our result means that quantum mechanics is the case where this freedom is maximal given a set of reasonable axioms.  

\section{How precedence builds up}

The action of the principle of precedence has to be restricted to cases in which the number of precedents of a measurement is large.  Otherwise, the first result with no precedence would be chosen randomly and that result would be the sole precedent for the second result, which would imply that all future measurements would repeat the first random choice.  To avoid this a different principle is required while the precedents build up an ensemble which fills out the elements of a density matrix.  One intriguing suggestion has been made by Markus Muller\cite{Markus-unpublished} who proposes that nature tries to induct from the first randomly chosen results the simplest possible rule.  This can be understood as saying that it is more efficient for nature to store a simple rule to generate the ensemble of outcomes than it is to store the whole ensemble of outcomes itself.  This leads to a hypothesis that nature chooses the simplest rule, in the sense of algorithmic information theory, which accounts for the first small number of precedents\footnote{Adrian Kent has proposed a related but different idea that theorists might choose theories on the basis of having the simplest expression in terms of algorithmic information theory\cite{AdrianK}.}.  Such a principle of a simplest rule contrasts interestingly with the principle that the state requires as much information as possible to specify.

This suggests that the laws of nature are the result of a minimalization, not of an action, but of the information  needed to express a rule that propagates future cases from past cases.  So rather than a principle of least action we will formulate dynamics as a principle of least information.  

There is of course a difficulty with the idea that the laws of nature are as simple as possible, which is that they aren't.  Free theories are simpler than interacting theories because it takes less information to specify linear laws then it does non-linear laws.  But a linear world would have no interactions and so no relations to define properties of subsystems.  So perhaps the principle that the world is relational forces it to be interacting.  Then we require the simplest possible non-linear law.  That is easy to satisfy: the simplest non-linear equations are quadratic.  Remarkably, both general relativity and Yang-Mills theory can be expressed as quadratic equations through Plebanski's trick of adding auxiliary variables\cite{universal,unify}. This basically follows from the fact that gauge couplings modify linear field equations by the addition of a quadratic term coupling matter with the gauge field.   Further, it is possible to conjecture that there may be a universal expression which unifies interacting gauge and gravitational theories, in which the fundamental dynamics expressed by a quadratic law\cite{universal}.  We can also consider the hypothesis that the state and the dynamical law are unified at a fundamental level, in such a way that the distinction between them is emergent and approximate\cite{unify}.  From this perspective the law that must be minimal is the one that evolves the state and dynamics together, as discussed in \cite{unify}.  This will be developed elsewhere.

\section{Discussion}

This proposal shows that, at the very least, we do not need to postulate timeless laws of nature to explain the success of physics as a predictive science. A weaker notion in which laws evolve through the accumulation of precedence suffices\footnote{Another approach to evolving laws is described in
\cite{Evolvinglaws}.}. The important thing is that this idea is testable, by the construction and study of entangled quantum states which are novel in the sense that they can reasonably be presumed not to have been produced in the history of the universe.  If one can construct such states,  one can study their evolution and possibly observe the evolution of novel precedents.

One can ask whether the freedom quantum systems have described here and in \cite{CK} means that hidden variables theories are impossible?  The answer is no, nothing could contradict the possibility of non-local, context dependent, hidden variables theories because several examples already exist, such as deBrogle-Bohm and \cite{real1}.  We have been concerned here with quantum mechanics as a description of small subsystems of the universe.  There could very well be a non-quantum theory that describes the whole universe, truncations of which to small subsystems yield quantum mechanics.  The freedom attributed to quantum systems could then be understood as being determined by information about the relations of subsystems to the whole which is lost when one truncates the cosmological theory to extract the behaviour of small subsystems.  

Someone might object that this interpretation of quantum theory takes the notion of copy, or similar preparation or measurement, as a primitive.  This is true.  But one should not have to apologize for the use of such primitives in an operational approach to quantum mechanics which, otherwise, takes notions of measurement to be primitives.  One might however, still ask how a system knows what its precedents are?  This is like asking how an elementary particle  knows which laws of nature apply to it. The postulate that general timeless laws act on systems as they evolve in time requires a certain set of metaphysical presuppositions. The hypothesis given here, that instead systems evolve by copying the responses of precedents in their past,  requires a different set of metaphysical presuppositions.  Either set of presuppositions can appear strange, or natural, depending on one's metaphysical preconceptions.  The only scientific question is which sets of metaphysical preconceptions lead to hypotheses which are confirmed by experiment.  

An important part of any package of metaphysical presuppositions is the relationship of laws of nature to time.  A timeless law cannot refer explicitly to the present or past, because those are thought to be subjective distinctions.  The formulation of quantum mechanics proposed here refers explicitly to the past and present and so only makes sense within a framework in which the distinctions between past, present and future are held to be real and objective.  This makes it possible to discuss objectively notions of laws evolving in time.  

This interpretation of quantum theory falls into a larger program of research  outlined in work with Roberto Mangabeira Unger\cite{RobertoLee}  The principles of this program are:
\begin{itemize}

\item{}The reality of time, so that notions such as present, past and future can be meaningfully used and distinguished in the fundamental laws of physics.

\item{}The uniqueness of a single universe.

\item{}The lack of any exact mirror or completely faithful copy of the universe, which extends to the absence of a completely faithful isomorphism to any mathematical object.

\end{itemize}

This proposal resolves a puzzle about cosmology which is what the laws of nature were doing before the big bang.  Were the laws waiting an infinite time for the universe to come into being so they would have something to apply to?   According to the idea presented here, without a universe there are no laws,  because the only law is precedence.   

Finally, some readers will ask whether there are any implications for whether human beings or animals have freedom to make choices not completely determined by the past.  This might arise by the generation of novel entangled states in neural processes.   Of course, allowing the possibility for novelty is not sufficient.  What would be necessary to realize the idea would be to discover that  the outcomes of neural processes are influenced by quantum dynamics of large molecules with entangled states, so that the lack of determinism of quantum processes is reflected in human choices and actions.  This could very easily fail to be the case.  Resolving this kind of question remains a goal for the distant future.

\section*{ACKNOWLEDGEMENTS}

The expression of the idea of precedence developed here owes a great deal to the work of Hardy and 
Masanes and Muller, after which my main result is a trivial step.  I am especially grateful to Lucien Hardy and Markus Muller for taking the time to explain their work as well as for comments on the manuscript.    
I am grateful also to Sabine Hossenfelder, Adrian Kent and Jaron Lanier for correspondence, conversation, comments on the manuscript and encouragement.  This work evolved out of a larger project with Roberto Mangabeira Unger, to whom I am grateful for a greatly stimulating collaboration.  
Research at Perimeter Institute
for Theoretical Physics is supported in part by the Government of
Canada through NSERC and by the Province of Ontario through MRI.

\end{document}